\documentstyle[pra,aps,epsfig]{revtex}
\begin{document}
\draft
\title{Quasi-static Ionization of Rydberg Alkali-metal Atoms:\\ a classical view of the $n^{-5}$ scaling}
\author{Luca Perotti}
\address{Center for Nonlinear and Complex Systems, Universit\'a degli studi dell'Insubria, Via Valleggio 11, Como 22100, Italy}
\date{\today}
\maketitle

\begin{abstract}
{\bf Abstract:} A fully classical explanation of the nonhydrogenic ionization threshold for low angular momentum Rydberg states of Alkali-metal atoms in a linearly polarized low frequency monochromatic microwave field is given: the classical equivalent to the quantum rate-limiting step, which is responsible for the $n^{-5}$ scaling and which according to the literature initiates what then continues as essentially classical diffusion, is found.
\end{abstract}

\pacs{32.80.Rm, 05.45.Mt, 72.15.Rn}

\section{Introduction}

Laboratory experiments \cite{gal,gal2} have shown that while the ionization threshold of high angular momentum Rydberg Alkali-metal atoms in a linearly polarized monochromatic microwave field with intensity $F$ and frequency $\omega \ll n_0^{-3}$, where $n_0$ is the initial principal quantum number, follows the hydrogenic behaviour $F_{th}\simeq(16n_0^{4})^{-1}$, the threshold for low angular momentum states behaves as $(3n_0^{5})^{-1}$. The quantum numerical simulations of A. Buchleitner and A. Krug \cite{buc,buc1,krug} have substantially confirmed this picture, even though they were not able to definitively rule out for low angular momentum states a $F_{th}\sim (90n_0^{4})^{-1}$ behaviour \cite{krug} for which on the other hand no theoretical explanation is known.

In the literature it is often found the opinion that this $(3n_0^{5})^{-1}$ threshold behaviour is nonclassical: see e.g. Ref. \cite{gal3} where the process is described as an essentially classical diffusion one {\it initiated} by a rate-limiting step which is a Landau-Zener transition at a quantum avoided crossing. On the other hand we are in a regime where classical behaviour can be expected on the relatively short time scales of the experiments: ionization of high angular momentum states follows the {\it classical} hydrogenic behaviour $F_{th}\simeq(16n_0^{4})^{-1}$; dynamical quantum localization does has no relevance at such low frequencies as the quantum delocalization border is below the experimental ionization threshold \cite{chi1,lu}; tunneling \cite{dam} and multiphoton processes happen at much longer time scales; and finally avoided crossings -the quantum resonances which are the foundation of the quantum explanation given for the $(3n_0^{5})^{-1}$ threshold- are in most cases clearly related to easily identifiable classical resonances, the magnitude of the splitting being proportional to the width of the classical resonance zone \cite{tab}. The aim of the present paper is to show that a fully classical explanation of the $(3n_0^{5})^{-1}$ threshold exists and thus add to the evidence in favour of such a scaling: on the energy surface of a Rydberg Alkali-metal atom in a static electric field the atomic core potential induces resonances; their overlap -which we shall show happens at a field intensity $F \approx (3n_0^{5})^{-1}$- brings chaos \cite{chi} and if the system is chaotic, a slowly varying field cannot be followed adiabatically \cite{lan,arn} by the system and energy diffusion will ensue.

The paper is thus organized: section \ref{due} summarizes the quantum explanation found in the literature; in section \ref{tre} classical numerical simulations are presented which suggest that a classical explanation is also possible; in section \ref{qua} the numerical results of section \ref{tre} are explained through a study of the phase space structures on the constant energy surfaces of an alkali-metal atom in a static electric field. Finally in section \ref{cin} the results of section \ref{qua} are applied to the case of a slowly oscillating harmonic field to derive the experimentally observed ionization threshold. Section \ref{sei} summarizes the results obtained.

\section{The Quantum Picture}\label{due}

The standard quantum explanation of the $(3n_0^{5})^{-1}$ behaviour is given \cite{gal3,gal4} in terms of core-induced Landau-Zener interactions \cite{laze} between states of Stark manifolds with principal quantum number differing by $1$ and field-induced Demkov-like interactions \cite{dem} within each manifold, as exemplified in the qualitative plot shown in Figure \ref{fig1}, which shows the optimal ionization path for a $n_0 \gg 1$ s-state as the electric field slowly oscillates at the threshold amplitude $F =(3n_0^{5})^{-1}$. Interactions which can be approximated by the Landau-Zener model (characterized by field dependent diagonal matrix elements and constant off-diagonal ones) are marked by circles; interactions which instead can be approximated by a Demkov-like model (characterized by constant diagonal matrix elements and field dependent off-diagonal ones) are marked by squares. As the field rises the $n_0,s$ state first undergoes an avoided crossing of the Landau-Zener type with the highest energy state of the $(n_0-1), m=0$ manifold and then, at $F=(3n_0^{5})^{-1}$, with the lowest energy state of the $n_0, m=0$ manifold; when the field decreases the population distributes among these three states and -as the field approaches zero- among all the states of the two manifolds, due to Demkov-like interactions. When the field rises again, the peak field is not high enough to couple the $(n_0-1), m=0$ manifold to the $(n_0-2), m=0$ one; on the other hand several states of the $n_0, m=0$ get to be coupled to the $(n_0+1), m=0$ manifold, so that diffusion toward higher energies and eventually ionization ensues. The actual ionization rate is not easily evaluated as the population fraction exchanged at each avoided crossing depends in a nontrivial manner on the oscillation frequency of the field and its peak intensity \cite{stu}.

Since, in the experiments performed up to now, high angular momentum states are not excited selectively in the quantum azimuthal number $m$, only a small fraction of their population is initially in the $m=0,\pm 1$ Stark manifolds, which are those displaying significant core-induced avoided crossings \cite{gal3}, while most of the population is in the high $m$ manifolds where the avoided crossings are too narrow to induce nonhydrogenic behaviours. Manifolds with the same principal quantum number $n$ but different $m$ are not coupled by the electric field; the population of high angular momentum states is therefore trapped in the high $m$ manifolds and the states themselves display a hydrogenic ionization threshold $F_{th}\simeq(16n_0^{4})^{-1}$.

\section{Preliminary Classical Exploration}\label{tre}

As we have seen in the quantum description above, because the electric field varies very slowly, most of the evolution is adiabatic, and the results obtained for Alkali-metal atoms in a static electric field can be used to analyze the case of a slowly varying field. Our first step will therefore be a classical study of the static field case. From the discussion in the above section  it is also clear that the case $m = 0$ is exactly the case where the nonhydrogenic behaviour we want to investigate appears most clearly; I have therefore chosen to restrict my study to such a case.

The motion of the perturbed system being restricted to a plane, I consider a two dimensional Alkali-metal atom model \cite{co} in the $\{x,z\}$ plane:  
\begin{eqnarray}
H_0 = {1\over 2}\left(p_x^2+ p_z^2 \right) - {1\over {r}} - {{\beta e^{-\alpha r}}\over {r}},\label{h0}\\
r=\sqrt{x^2+z^2}\nonumber
\end{eqnarray}
where $p_x$ and $p_z$ are the conjugate momenta to the spatial coordinates $x$ and $z$. The first two terms in eq. (\ref{h0}) represent the Hamiltonian for a Hydrogen atom with zero $z$ component of the angular momentum $L_z = 0$ (that is: azimuthal quantum number $m = 0$) and the last term is the simplest known model for the nonhydrogenic core potential, $\beta$ being the core charge and $\alpha$ a parameter proportional to the inverse of the core radius \cite{no1}. The values of the parameters which give the best fit to lithium have been found to be $\alpha=2.13$ and $\beta=2$ \cite{co}.

We now add to the free atom Hamiltonian (\ref{h0}) a static field potential $V = F z$; the system described by the Hamiltonian $H=H_0 +V$ is invariant under the scaling $x \rightarrow x/ n_0^2$, $z \rightarrow z/ n_0^2$, $p_x \rightarrow p_x n_0$, $p_z \rightarrow p_z n_0$, $H \rightarrow H n_0^2$, $t \rightarrow t/ n_0^3$, $F \rightarrow F n_0^4$, $\alpha \rightarrow \alpha n_0^2$, $\beta \rightarrow \beta$. 
Such a scaling would result in a hydrogenic threshold behaviour as can be easily seen from the scaling for the field intensity $F$; on the other hand the core parameter $\alpha$ is fixed for each atom type and therefore cannot be changed. Changing $n_0$ without changing $\alpha$ means changing the relative dimension of the core to the orbit average radius and results in a non-hydrogenic threshold behaviour as can be seen from the following preliminary numerical results, presented -in accordance with the current literature- in semiparabolic coordinates scaled to the initial quantum number $n_0$:
\begin{eqnarray}
u={{\sqrt{r+z}}\over {n_0}}\\
v={{\sqrt{r-z}}\over {n_0}},
\end{eqnarray}
and in their canonically conjugate momenta $p_u$ and $p_v$.

My findings are summarized in Figs. (\ref{fig2}) and (\ref{fig3}) where single orbit Poincar\'e surfaces of section (SOS) for $n_0=40$ and $n_0=320$ are shown. A SOS for a system orbit is given by the points where the orbit crosses a given plane with positive (negative) velocity. Those shown are for the $(v, p_v)$ plane at $u=0$ with positive velocity ${\dot u} = du/dt > 0$. A comparison of the right side of each figure, for which $F n_0^5= 0.04$, with the left side where $F n_0^5= 0.32$ clearly illustrates that the fraction of the energy surface a single orbit is able to explore varies as $F n_0^5$; in particular it can explore the entire surface when $F=(3n_0^{5})^{-1}$, as can be seen from the left hand side of both figures. The form taken by this exploration is on the other hand different in the two cases: outside of the central region, where the chaotic behaviour first becomes evident \cite{co}, the $n_0=320$ orbit appears much more regular than the $n_0=40$ one; this is a consequence of the fact that far from the nucleus the system is hydrogen-like and therefore scales as $F n_0^4$.

Our aim is to now understand why this happens; we shall do it by applying Chirikov's resonance overlap criterion for transition to chaos \cite{chi} in a form suitable to autonomous systems \cite{esc}.

\section{The Constant Field Classical Model}\label{qua}

The quantum mechanical studies quoted in section \ref{due} have shown that the relevant feature in the quasistatic ionization of alkali-metal atoms is the core-induced avoided crossing between the highest lying state of the $n_0$ Stark manifold with the lowest one of the $n_0 +1 $ one (in the hydrogen atom, the levels cross because of symmetry \cite{klep}). This avoided crossing takes place at an electric field strong enough that the electric field potential dominates over the core potential; it is therefore convenient to consider an (hydrogen-like) atom without core potential in a static electric field as our unperturbed system and the core itself as our perturbation; this puts us out of the range of the weak electric field approximation used in Refs. \cite{perc,lev}. 

To apply Chirikov's criterion it is convenient to write the Hamiltonian in action-angle variables, so that we can explicitly calculate the characteristic frequencies of the unperturbed system. To do this we first write the Hamiltonian of the unperturbed system (a two-dimensional hydrogen atom in a static electric field) 
\begin{equation}
H_h = {1\over 2}\left(p_x^2+ p_z^2 \right) - {1\over {\sqrt{x^2+z^2}}} + F z
\end{equation}
in parabolic coordinates $\{\xi,\eta\}$, defined by
\begin{eqnarray}
r={{\xi + \eta}\over 2}\label{x}\\
z={{\xi - \eta}\over 2},\label{z}
\end{eqnarray}
and their conjugate momenta $p_{\xi}$ and $p_{\eta}$. The resulting Hamiltonian
\begin{equation}
H_h = 2{{\xi p_{\xi}^2+\eta p_{\eta}^2}\over {\xi+\eta}}- {2\over {\xi+\eta}} + F {{\xi - \eta}\over 2}\label{hh}
\end{equation}
separates \cite{lan} and is therefore regular (non-chaotic). We can now pass to the action-angle variables $\{I,I_1,\lambda,\mu\}$ defined by the equations
\begin{eqnarray}
\xi = 2 I I_1 (1-\sin \chi_1) \label{xi}\\
\eta = 2 I (I-I_1) (1-\sin \chi_2) \label{eta}\\
p_{\xi} = {1\over {2 I}} {{\cos{\chi_1}}\over{1-\sin \chi_1}} \\
p_{\eta} = {1\over {2 I}}{{\cos{\chi_2}}\over{1-\sin \chi_2}} ,
\end{eqnarray}
where $I_1 \le I$, and $\chi_1$ and $\chi_2$ are auxiliary angles defined by 
\begin{eqnarray}
\lambda=-{{I_1}\over I}\cos \chi_1-{{I-I_1}\over I}\cos \chi_2 - \chi_2 +{{\pi}\over 2} \label{la}\\
\mu=\chi_2 - \chi_1 \label{mu}.
\end{eqnarray}
The two actions $I$ and $I_1$ are the classical equivalent of the parabolic quantum numbers $n$ and $n_1$ used in quantum mechanics to describe the interaction of a hydrogen atom with a static electric field \cite{lan1} and $\lambda$ and $\mu$ are their respective canonical angles.

In these coordinates, the Hamiltonian (\ref{hh}) reads, in first (linear) approximation, \cite{lan}
\begin{equation}
H = -{1 \over {2 I^2}}+ {3 \over 2} F I (2I_1 -I).\label{hh2}
\label{star}
\end{equation} 

We have considered only the linear term of the electric potential as the relevant crossing in the corresponding quantum system takes place in the linear regime of the Stark potential not only for the high angular momentum states but also for the low angular momentum ones which, due to the quantum defect, display at low electric fields only a quadratic Stark shift \cite{zim,note5}.

We can now add the core potential to the above Hamiltonian; to introduce it as a perturbation, we take its Fourier expansion:
\begin{eqnarray}
V =A_{0,0}(I, I_1)+ \Sigma_{k,k_1>0}2A_{k,k_1}(I, I_1)\cos{(k\lambda+k_1\mu)}.\label{fou2}
\end{eqnarray}
where the coefficients $A_{k,k_1}(I, I_1)$ are the semiclassical matrix elements of the perturbation \cite{lan1}. For the resonant terms we are interested in, we have -as we shall shortly see- $k \ll k_1$ and $k \ll \alpha I^2$; as long as $I_1$ is not too close to $0$ or $I$ (the condition $I_1,(I-I_1)>k_1^2 /(2\alpha I)$ must be verified), we can therefore write these coefficients as: 
\begin{eqnarray}
A_{k,-\bar{k}_1}(I, \bar{I}_1) =\beta {{e^{-\alpha I^2}}\over {I^2}}\left({{\alpha I^2+k}\over{\alpha I^2-k}}\right)^{k/2}I_{k_1}\left(\bar{I}_1\sqrt{\alpha^2 I^4-k^2}\right)I_{k-k_1}\left((1-\bar{I}_1)\sqrt{\alpha^2 I^4-k^2}\right)\label{a3}\\
\simeq {{\beta}\over {2\pi I^4 \alpha \bar{I}_1(1-\bar{I}_1)}} \exp{[-{{\bar{k}_1^2}\over{2\alpha}}{1 \over{\bar{I}_1(1-\bar{I}_1)}}]}\equiv {{\bar{A}_{k,-\bar{k}_1}(\bar{I}_1)}\over {I^4}}\label{aaaa},
\end{eqnarray}
where the $I_j(y)$'s are modified Bessel functions and we have introduced the scaled action $\bar{I}_1=I_1/I$ and the scaled index $\bar{k}_1 =-k_1/I$. The approximations made to derive eq. (\ref{aaaa}) are discussed in the appendix.

The scaled coefficients $\bar{A}_{k,-\bar{k}_1}(\bar{I}_1)$ appear therefore almost independent from $I$ and $k$. Figure (\ref{fig4}) shows the dependence on $\bar{k}_1$ of the exact expression for $\bar{A}_{1,-\bar{k}_1}(\bar{I}_1)$ eq. (\ref{a3}) for several values of $\bar{I}_1$ for lithium ($\alpha=2.13$ and $\beta=2$): the matrix elements are significant only for $\bar{k}_1\sim <1$. 

Figure (\ref{fig6}) shows instead the dependence of $\bar{A}_{1,-1}(\bar{I}_1)$ on $\bar{I}_1$, again using the xact expression eq. (\ref{a3}): there is a wide peak around the central value $\bar{I}_1=1/2$ where from eq. (\ref{aaaa}) we have 
\begin{equation} 
\bar{A}_{1,-1}(1/2) \simeq{{\beta}\over{\pi \alpha}}e^{-{2}\over {\alpha}}=\simeq 0.116,
\end{equation}
and drops to zero for $\bar{I}_1=0$ and $\bar{I}_1=1$.

Resonances induced by the core will take place when the phase of one of the terms of the Fourier expansion (\ref{fou2}) is stationary; this happens when the two unperturbed frequencies of motio for $\lambda$ and $\mu$,
\begin{eqnarray}
\omega_0\equiv {{\partial H}\over{\partial I}} = {1 \over {I^3}}+ 3 F (I_1 -I)\label{o1}\\
\omega_1\equiv {{\partial H}\over{\partial I_1}} = 3 F I \label{o2},
\end{eqnarray}
satisfy the resonance condition
\begin{equation} 
k \omega_0+ k_1 \omega_1 =0\label{res},
\end{equation}
with $k$ and $k_1$ two integers.

Since, for $F$ below the hydrogen ionization threshold, we have $\omega_0\gg \omega_1$, the most important resonances (those having large stationary terms) will be for $k= \pm 1$, so that $k_1$ is not too big; in particular from Figure (\ref{fig4}) we see that the matrix elements $A_{1,-k_1}(I, I_1)$ are significantly large only for $|k_1| \sim < I$. Substituting $k=1$ and imposing $-k_1 < I$ in the resonance condition eq. (\ref{res}), we obtain
\begin{equation} 
F > {1 \over{3I^5\left[1+{1\over {I^2}}\left(I-I_1\right)\right]}}. \label{sogl}
\end{equation}
For the high values of the principal action $I$ we are considering the second term in the denominator is negligible; eq. (\ref{sogl}) therefore becomes independent of $I_1$ and, since $I$ is the classical analogue of the principal quantum number $n_0$, corresponds to the experimental ionization threshold $F=(3n_0^{5})^{-1}$.

Obviously, the above argument only tells us that classical resonances are noticeably big only for $F>\sim (3n_0^{5})^{-1}$; whether they do overlap and thus generate global chaos on the energy surface is still an open question. To answer this question we have first to find the positions of the resonances on the energy surface. Let us therefore place ourself on a constant energy curve in $\{I,I_1\}$:
\begin{equation} 
E_0=-{1 \over {2I^2}}+{3 \over 2}FI(2I_1 -I);\label{sur}
\end{equation}
introducing the average action $I_0=1/ \sqrt{-2E_0}$ (corresponding to $I_1=I/2$), and the scaled quantities $\bar{I}= I/I_0$ and $F_0=F I_0^4$, eq. ({\ref{sur}) now reads
\begin{equation} 
3F_0\bar{I}^4(2\bar{I}_1-1)+\bar{I}^2-1=0.\label{sur2}
\end{equation}
In the same variables the resonance condition eq. (\ref{res}) for $k=1$ reads:
\begin{equation} 
3F_0\bar{I}^4(-k_1+1-\bar{I}_1)=1.\label{res2}
\end{equation}
If we now eliminate $\bar{I}$ between eq. (\ref{sur2}) and eq. (\ref{res2}), we obtain the condition
\begin{equation} 
3F_0 (-k_1+2-3\bar{I}_1)^2=(-k_1+1-\bar{I}_1). 
\end{equation}
which, solved for $k_1$, gives us
\begin{equation} 
-k_1^{\pm}= {{1-6F_0(2-3\bar{I}_1) \pm \sqrt{1-12F_0(1-2\bar{I}_1)}}\over {6F_0}}. 
\end{equation}
Only the positive root is of interest to us and for $12F_0 \ll 1$ we can expand the square root obtaining
\begin{equation} 
-k_1^{+}= {1 \over {3F_0}}+5\bar{I}_1-3 -3F_0(1-2\bar{I}_1)^2 +......\simeq {1 \over {3F_0}}+5\bar{I}_1-3\label{kap} 
\end{equation}

Since $\bar{I}_1\in [0,1]$, we have five resonances (six if $1/(3F_0)$ is an integer) at equally spaced values of $\bar{I}_1$ between $0$ and $1$: 
\begin{equation} 
\bar{I}_1= {{J-\left.{1 \over {3F_0}}\right|_{mod 1}}\over 5},\hspace{0.5in}J=(0),1,...5.\label{n1}
\end{equation}
where the value $0$ of the index $J$ has been put in parenthesis to indicate it is possible only for $1/(3F_0)$ an integer. The corresponding indices $k_1$ will read
\begin{equation} 
-k_1= \left[{1 \over {3F_0}}\right]+J-3 \label{kap1} 
\end{equation}
where $[.]$ denotes integer part.
The resonance values for $\bar{I}$ will instead be
\begin{equation} 
\bar{I} = {1 \over {\left\{1+3F_0\left[{4\over 3}\left({J-\left.{1 \over {3F_0}}\right|_{mod 1}}\right)-2\right]\right\}^{1/4}}},\label{act1}
\end{equation}
which, since we are considering $3F_0 \simeq 1/I_0 \ll 1$, we can approximate with $1$.

From Figure (\ref{fig6}) it is clear that the widths of the first and last resonances will be zero (for $1/(3F_0)$ an integer) or close to zero, while the other four will be of comparable widths. We now want to compare the average width of these resonances to the separation of two consecutive resonances, if the former is larger than the latter, the resonances overlap and we have chaos on the energy surface (Chirikov's resonance overlap criterion \cite{chi}).

To apply Chirikov's resonance overlap criterion for transition to chaos to autonomous systems, we have to take into account energy conservation \cite{esc}: the resonance width to be considered is the one along the constant energy curve in $\{I,I_1\}$ containing the resonance center itself; from eq. (\ref{star}) the versor ${\bar r}$ tangent to the energy curve reads $\bar r = \{\omega_1/ |{\bar \omega}|, -\omega_0/ |{\bar \omega}|\}$ where $\omega_0$ and $\omega_1$ are given by eqs. (\ref{o1},\ref{o2}) and $|{\bar \omega}|=\sqrt{\omega_0^2+\omega_1^2}$. Following Ref. \cite{esc}, the resonant Hamiltonian in the restricted phase space of the tangent action $J= [(I-I^{(r)})\omega_1 -(I_1-I_1^{(r)})\omega_0]/ |{\bar \omega}|$ and its conjugate angle $\varphi= (\lambda\omega_1 -\mu\omega_0)/ |{\bar \omega}| $ where $I^{(r)}$ and $I_1^{(r)}$ are the actions at the $(k,k_1)$ resonance, reads: 
\begin{eqnarray} 
H_r={1\over 2} a J^2 + {2A_{k,k_1}(I^{(r)}, I_1^{(r)})}\cos{(\Omega \varphi )},\label{ar}\\
\Omega= k|{\bar \omega}|/\omega_1,\\
a={\bar r} {\bar {\bar \sigma}} {\bar r},
\end{eqnarray}
where $\bar {\bar \sigma}$ is the matrix of the second derivatives of $H$ with respect to $I$ and $I_1$. As $|{\bar \omega}| \simeq \omega_0$, we have $\Omega \simeq |k_1|$ and $|k_1|$ is the number of islands in the chain along the variable $\varphi$.

The half width of the $(k,k_1)$ resonance in $J$ is therefore:
\begin{eqnarray} 
W=2\sqrt{{2A_{k,k_1}(I^{(r)}, I_1^{(r)})}\over |a|},\label{wi1}
\end{eqnarray}
and since $|{\bar \omega}| \simeq \omega_0$, it is also the width in $I_1$, while the width in $I$ is negligible.

Since $F_0 \ll 1$ we have, dropping the index ${(r)}$ to avoid too cumbersome a notation,
\begin{equation} 
a=\left({3 F I }\over {|{\bar \omega}|}\right)^2 \left({{-5}\over {I^4}} +3 F - 6 F {{I_1}\over I}\right)= \left({{3F_0\bar{I}^2}\over {I_0^2}}\right)^2 {{-5+3F_0\bar{I}^4(1-2\bar{I}_1)}\over{[1+3F_0\bar{I}^4(\bar{I}_1-1)]^2+(3F_0\bar{I}^4)^2}}\simeq -5 \left({{3F_0\bar{I}^2}\over {I_0^2}}\right)^2.\label{aa1}
\end{equation}

Taking for $k_1$ the approximate average value from eq. (\ref{kap}) $k_1=-1/(3F_0)$ and $k=1$, so that $k-k_1 \simeq -k_1$; taking also $I_1 = I/2$ (this latter being the value for which $A_{k,k_1}(I, I_1)$ is maximum, it will give an overestimate of the width) we obtain, using eq. (\ref{aaaa}) and remembering that for $I_1 = I/2$ we have $\bar{I}=1$,
\begin{equation} 
A_{1,-{1\over{3F_0}}}(I_0,I_0/2) \simeq{{\beta}\over{\pi \alpha I_0^4}}e^{-{2}\over {\alpha(3F_0 I_0)^2}}. 
\end{equation}
The maximum resonance half width on the energy surface, scaled to the average principal action $I_0$, will therefore be 
\begin{equation} 
{W\over{I_0}}\simeq {2 \over {3F_0 I_0}}\sqrt{{2\beta}\over{5\pi \alpha}}e^{-{1}\over {\alpha(3F_0 I_0)^2}}. \label{wid}
\end{equation}
Figure (\ref{fig7}) shows $W/I_0$ as a function of $3F_0 I_0$ for the lithium parameters. The dotted curve is eq. (\ref{wid}); the full curves are instead the exact scaled widths calculated at $I_0=10$ for (from top to bottom) $\bar{I}_1=0.50$, $0.20$, $0.10$, and $0.05$. The first overlap will happen when the $\bar{I}_1=0.5$ width becomes larger than $0.10$, that is for $3F_0 I_0 \simeq 0.4$; to have global chaos on the energy surface we instead need the extreme resonances to both overlap with their nearest neighbour and to reach the extremes in $I_1$ of the energy manifold; this happens when the $\bar{I}_1=0.20$ width becomes larger than $0.20$, that is at $3F_0 I_0 \simeq 0.7$, in reasonable agreement with the numerical results from section \ref{tre}: the deviation is within the usual factor $2$ expected for Chirikov's criterion. Worth noting is the fact that, as the perturbation is the atomic core, the resonance width depends on the electric field $F$ only indirectly, through the resonance index $k_1$ (see eq. (\ref{kap1})) and the energy curve shape. The first dependence gives the growth of the resonance width, the second one its slow decrease at high $F$ where on the other hand the linear approximation for the electric field potential used in the Hamiltonian (\ref{hh2}) is no more valid. 

While eq. (\ref{wid}) shows that the average behaviour depends on the product $3F_0I_0$ only, the details depend on $3F_0$ and $I_0$ separately: using eqs. (\ref{aaaa}), (\ref{n1}), (\ref{kap1}), (\ref{wi1}), and (\ref{aa1}), we obtain, remembering that from eq. (\ref{act1}) $I \simeq I_0$,
\begin{equation} 
{W\over{I_0}}\simeq {2 \over {3F_0 I_0}}\sqrt{{\beta}\over{\pi \alpha \sqrt{(J-\left.{1 \over {3F_0}}\right|_{mod 1}) (5-J+\left.{1 \over {3F_0}}\right|_{mod 1})}}}e^{-{{1}\over {\alpha}} \left({{\left[{1 \over {3F_0}}\right]+J-3}\over {2 I_0}}\right)^2 {1 \over {(J-\left.{1 \over {3F_0}}\right|_{mod 1}) (5-J+\left.{1 \over {3F_0}}\right|_{mod 1})} }}.  \label{wid2} 
\end{equation}

Fig. (\ref{fig8}) shows the positions (eq. (\ref{n1})) and widths (eq. (\ref{wid2})) of the resonances in $\bar{I}_1=I_1/I\simeq I_1/I_0$ as a function of $3F_0I_0$ for a) $I_0=10$ and b) $I_0=60$: in both cases the general behaviour is the same: the first overlaps happen for $\bar{I}_1=0.5$ at $3F_0 I_0 \simeq 0.41$ (see the detail in Fig. (\ref{fig9})) and the entire $\bar{I}_1$ range is covered by the resonances for $3F_0 I_0 \simeq 0.7$, but the vertical symmetry is much better for $I_0=60$; this is easily explained by noting that, because of eq.(\ref{kap}), the speed of motion of a resonance toward higher $\bar{I}_1$ values with increasing $3F_0 I_0$ increases with $I_0$, and that in our approximation the resonance width depends from $I_1$ only through $A_{k,k_1}(I,I_1)$ and is therefore symmetric in $\bar{I}_1$ and $1-\bar{I}_1$ {\it along the resonance curves} ($k_1=const$).

\section{The Variable Field Classical Model}\label{cin}

Let's consider a cyclical perturbation of a given system; an action is an (approximate) adiabatic invariant if the unperturbed frequency of motion of the associated angle variable is much larger than the fractional rate of change of the perturbation itself \cite{lan}; even though there is no change of energy over a whole cycle of the perturbation, the energy of the system changes during the time the perturbation is varied. The opposite limit is that of sudden or diabatic variation: in this case the energy of the system does not change at all during the time the perturbation is varied. 

In a few words: diabatic variation of a perturbation means that the motion of the unperturbed system is very slow compared to the rate of change of the perturbation and therefore a whole cycle of the perturbation brings no change to the system, which had no chance to evolve. Adiabatic variation means instead that the motion of the unperturbed system is very fast compared to the rate of change of the perturbation so that the system can follow the slowly varying orbit it is on and again a whole cycle of the perturbation brings no change to the system, {\it apart from a vector phase}, sum of a dynamical part (the integral over the pulse time of the istantaneous frequency for each action) and of a geometrical part (the so called Hannay angle \cite{han}). 

Different orbits having different periods, adiabaticity depends on initial conditions. Two classes of orbits are intrinsecally non-adiabatic. The first one is that of the seperatrices: orbits with {\it infinite period} which separate regions of phase space characterized by different kinds of motion, e.g. rotation and libration for a pendulum. The other class is that of chaotic orbits, beacause of exponential instability. 

The really interesting case is the case in between the two above: in this case motion of the unperturbed system and rate of change of the perturbation are comparable and the system neither stays still nor evolves along the varying orbit it was initially on. It instead moves to different static perturbation orbits so that in general energy exchange will ensue. 

A particularly interesting such case is when the variation of the perturbation causes the system to non-adiabatically cross over (either through a separatrix or a chaotic region) from a region of phase space to another where motion is of a different nature; in this case the phase acquired by the system during the adiabatic part of the evolution can play an essential function in determining the final energy (see e.g. \cite{die}).

In our case we slowly vary the electric field according to the sine law $F(t)=F\sin{(\omega_f t)}$ so that the fractional rate of change reads $\omega_f / \tan(\omega_f t)$; apart from the cases $\omega_f t=N\pi$, $N$ an integer, where it goes to zero, or $\omega_f t=(2N+1)\pi/2$ where it goes to infinity \cite{note3}, we can use the approximate value $\omega_f$.   

Considering as our unperturbed system a Hydrogen atom in a static electric field the frequencies to be considered would be those associated with the two independent actions $I_1$ and $I_2$: in the electric field linear regime, they give the conditions
\begin{eqnarray} 
\omega_f  \ll {1 \over {I^3}} + 3FI_1 = \omega_0 + \omega_1\\
\omega_f  \ll {1 \over {I^3}} - 3FI_2 = \omega_0;
\end{eqnarray}
which, as in our case $FI^3 I_i < FI^4 \ll 1$, reduce in both cases to
\begin{equation} 
\omega_f I^3 \ll 1.
\end{equation}
The above condition is the exact definition of the regime we are interested in and is therefore always verified.

In the case the unperturbed system is instead an Alkali-metal atom in a static electric field, we shall distingush two regimes: low electric fields $F \ll (3n_0^{5})^{-1}$; and threshold electric fields $F \simeq (3n_0^{5})^{-1}$, where, as chaos is global, adiabatic evolution is impossible everywhere. 

For low electric fields the situation is more complicated, as the phase space is mixed: deformed tori and resonance islands coexist in it; far from the resonance islands the two actions $I_1$ and $I_2$ are still good approximations of the actual actions of the system and we still have adiabatic evolution for $\omega_f I^3 \ll 1$; in the regions on the border of the resonance islands the motion is instead chaotic and no adiabatic motion is possible \cite{note2}; finally, inside the islands the system is approximated by a pendulum one \cite{chi}, from eq. (\ref{ar}) we therefore have that, close to the center of an island, the frequency of motion around the center of the island itself is $\omega_i = \Omega \sqrt{2a A_{k,k_1}(I, I_1)}$, so that the adiabaticity condition reads:
\begin{equation} 
\omega_f I^3 \ll {{15}\over 2} F_0 {W \over I};
\end{equation}
using eq. (\ref{wid}) as an approximate evaluation of $W$ this condition becomes:
\begin{equation} 
\omega_f I^3 \ll {1 \over {I}}\sqrt{{10\beta}\over{\pi \alpha}}e^{-{1}\over {\alpha(3F_0 I)^2}}\simeq  {{\sqrt{3}} \over {I}} e^{-{1}\over {2(3F_0 I)^2}},\label{ada}
\end{equation}
where in the last expression we have introduced the parameters for Lithium. As $I\gg 1$, the condition is a very stringent one and is never verified for the range of parameters we are interested in; adiabaticity is therefore not possible also {\it inside} the resonance islands themselves.
 
Let us now see what happens when the electric field in our system is slowly increased: from Figs. (\ref{fig8}) and (\ref{fig9}) we see that the resonances appear at $I_1 = 0$ and {\it move} to higher values of $I_1$ \cite{note4} growing in size up to $I_1=I/2$ and then decreasising again. This might at first sight suggest as a ionization mechanism the trapping of orbits in the resonance islands which would then transport them and finally release them either when the resonance islands themselves break because of overlap with nearby resonances \cite{yuan} or when the island size decreases (either at high $I_1$ for increasing field, or at low $I_1$ for decreasing field). Such a mechanism could in principle work for $F < (3n_0^{5})^{-1}$; it would on the other hand require adiabatic (or quasi-adiabatic) evolution within the resonance islands, which as we have seen is far from the case for the range of parameters we are interested in: as the motion inside the resonance island is at its fastest still much slowler than the rate of field change, the islands themselves will not be seen by the orbits. 

The only possible mechanism is therefore diffusion in the chaotic sea, which happens only for $F \simeq (3n_0^{5})^{-1}$. As it was the case in the quantum description summarized in Sect. \ref{due}, an actual evaluation of the diffusion and ionization rates does not appear easy: the only data we have for comparison are the numerical ones in Ref. \cite{krug} and they show that over long interaction times the decay of the survival probability is neither exponential, as expected for a completely chaotic phase space \cite{expo}, nor algebraic, as expected in mixed phase space \cite{alge}. On the other hand, for short interaction times, while for the near-threshold plot in Fig. 6.8 from Ref. \cite{krug} the figure resolution is too low to extract any data for comparison, the high field plots seem to indicate an initial behaviour still in accordance with the classical decay time $t_D = \pi\omega_f^{4/3}/I_0F^2$ given in Ref. \cite{cas} for $\omega_f > I_0^{-3}$.

\section{Conclusions}\label{sei}

We have shown here that while the ionization of excited hydrogen atoms in a quasistatic monochromatic field is regular, that of low angular momentum alkali-metal atoms is chaotic. Core induced resonances overlap at $F \simeq (3n_0^{5})^{-1}$, thus triggering chaotic diffusion which results in ionization. The motion within the single resonance islands being extremely slow, the islands themselves are not seen by the system during the pulse and therefore allow no efficient path to ionization.

The present paper concludes my study in terms of nonlinear dynamics of the ionization mechanisms for excited Alkali-metal atoms in microwave fields initiated in Ref. \cite{lu} and prompted by the individuation of three different regimes in the numerical studies by A. Buchleitner and A. Krug \cite{buc,buc1,krug}: in the high frequency regime $\omega_f I^3 > 1$ the ionization threshold is determined by the same quantum localization mechanism as for excited Hydrogen atoms \cite{cas}; in the intermediate frequency regime $\omega_f I^3 \sim < 1$ quantum localization again detemines the ionization threshold, but the Alkali-metal atoms quantum defects raise the dimensionality of the problem \cite{lu}; finally, as we have shown in the present paper, in the low frequency regime $\omega_f I^3 \ll 1$, classical dynamics, which for excited Hydrogen atoms essentially determines the ionization threshold alredy for $\omega_f I^3 < 1$, determines the ionization threshold for excited Alkali-metal atoms too, which, due to core effects, is chaotic and much lower than the Hydrogen one. 

\section{Acknowledgements}

The author wishes to thank P. O'Mahony and G. Mantica for helpful discussions and suggestions and the University of London for their hospitality when working on this paper. The present work was conducted within the frame of the QTRANS (``Quantum Transport on an atomic Scale") european community network (contract number HPRNT-CT-2000-00156).

\appendix
\section{Equations of motion in semiparabolic coordinates}
\label{app:par}

We derive here the exact expression of the coefficients $A_{k,k_1}(I, I_1)$ of the Fourier expansion of the core potential and discuss the approximation eq. (\ref{aaaa}) given in the text: the coefficients Fourier expansion 
\begin{eqnarray}
V \equiv - {{\beta e^{-\alpha \sqrt{x^2+z^2}}}\over {\sqrt{x^2+z^2}}}= \Sigma_{k,k_1}A_{k,k_1}(I, I_1)e^{i(k\lambda+k_1\mu)}\label{fou}
\end{eqnarray}
are given by the integral
\begin{eqnarray}
A_{k,k_1}(I, I_1)= -{1\over{(2\pi)^2}}\int_0^{2\pi}{d\lambda}\int_0^{2\pi}{d\mu  {{\beta e^{-\alpha {\sqrt{x^2+z^2}}}}\over {\sqrt{x^2+z^2}}}e^{-i(k\lambda+k_1\mu)}}. 
\end{eqnarray}
To evaluate it, we change integration variables from $\lambda$ and $\mu$ to $\chi_1$ and $\chi_2$: from (\ref{la}) and (\ref{mu}) the Jacobian determinant of the transformation reads
\begin{equation} 
D(\chi_1 , \chi_2)= {{I_1}\over I}\sin \chi_1+{{I-I_1}\over I}\sin \chi_2-1;\label{jac}
\end{equation}
using then eqs. (\ref{x}), (\ref{z}), (\ref{xi}), and (\ref{eta}), we obtain \cite{tav}:
\begin{eqnarray}
A_{k,k_1}(I, I_1)=\hspace{5.8in}\nonumber\\
= (-i)^k {1\over{(2\pi)^2}}{{\beta e^{-\alpha I^2}}\over {I^2}}\int_0^{2\pi}{d\chi_1 e^{\alpha I I_1 \sin \chi_1 +i\left(k{{I_1}\over I}\cos \chi_1 +k_1 \chi_1\right)}}\int_0^{2\pi}{d\chi_2  e^{\alpha I (I-I_1) \sin \chi_2 +i\left[k{{I-I_1}\over I}\cos \chi_2 +(k-k_1) \chi_2\right]}}= \nonumber\\
=\beta {{e^{-\alpha I^2}}\over {I^2}}\left({{\alpha I^2+k}\over{\alpha I^2-k}}\right)^{k/2}I_{k_1}\left({{I_1}\over{I}}\sqrt{\alpha^2 I^4-k^2}\right)I_{k-k_1}\left({{I-I_1}\over{I}}\sqrt{\alpha^2 I^4-k^2}\right),\label{aa}
\end{eqnarray}
where the $I_j(y)$'s are modified Bessel functions.

Because for integer indices the modified Bessel functions have the property $I_{-j}(y)=I_j(y)$, it is clear from eq. (\ref{aa}) that $A_{-k,-k_1}(I, I_1)=A_{k,k_1}(I, I_1)$. Eq. (\ref{fou}) can therefore be written as:
\begin{eqnarray}
V =A_{0,0}(I, I_1)+ \Sigma_{k,k_1>0}2A_{k,k_1}(I, I_1)\cos{(k\lambda+k_1\mu)}.
\end{eqnarray}
which is the expression given in the text, eq. (\ref{fou2}).

If we now introduce the scaled action $\bar{I}_1=I_1/I$ in eq (\ref{aa}), we can easily see that, if $k \ll \alpha I^2$, the arguments of the modified Bessel functions scale approximately  as $I^2$. Moreover, whenever the index $j$ is small with respect to the argument $y$, we can use the expansion \cite{abr}
\begin{eqnarray}
I_j(y)={{e^y}\over {\sqrt{2\pi y}}}\left[1-{{4j^2-1}\over {8y}}+{{(4j^2-1)(4j^2-9)}\over {2!(8y)^2}}- {{(4j^2-1)(4j^2-9)(4j^2-25)}\over {3!(8y)^3}}+...\right]
\end{eqnarray}
which, if $j$ is itself large, can be further simplified as
\begin{eqnarray}
I_j(y)\simeq{{e^y}\over {\sqrt{2\pi y}}}\left[1-{{j^2}\over {2y}}+{{j^4}\over {2!(2y)^2}}- {{j^6}\over {3!(2y)^3}}\right] \simeq{{e^{y-{{j^2}\over {2y}}}}\over {\sqrt{2\pi y}}} .\label{ii}
\end{eqnarray}
Introducing the scaled index $\bar{k}_1 =-k_1/I$, assuming $k \ll k_1$ and $I_1,(I-I_1)>k_1^2 /(2\alpha I)$, and substituting eq. (\ref{ii}) in eq. (\ref{aa}), the matrix elements read
\begin{eqnarray}
A_{k,-\bar{k}_1}(I, \bar{I}_1) \simeq {{\beta}\over {2\pi I^4 \alpha \bar{I}_1(1-\bar{I}_1)}} \exp{[-{{\bar{k}_1^2}\over{2\alpha}}{1 \over{\bar{I}_1(1-\bar{I}_1)}}]},\label{aaa2}
\end{eqnarray}
which is the expression given in the text. From eq. (\ref{aaa2}) we see that $A_{k,-\bar{k}_1}(I, \bar{I}_1)\sim I^{-4}$; to give an idea of the residual dependence on $I$,  Figure (\ref{fig5}) compares the scaled matrix element calculated using eq. (\ref{aaa2}) with the exact ones for $I=10$ and $I=20$: even on a logarithmic scale the differences are small .

\begin{figure}[htbp]
\centering\epsfig{file=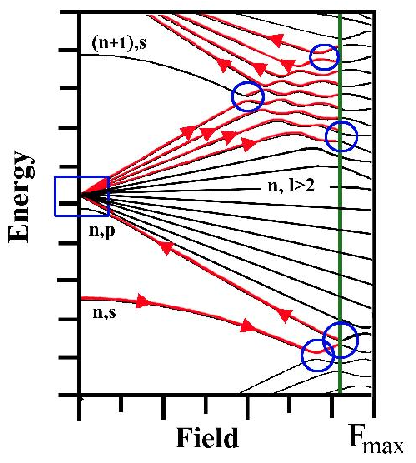,width=0.9\linewidth}
\caption{(Color online) The optimal ionization path for a $n_0 \gg 1$ s-state in a quasistatic microwave field oscillating between $\pm F_{max}\sim (3 n_0^5)^{-1}$. The rectangular box indicates the Demkov-like interaction region for the $n_0$ manifold; circles mark instead the Landau-Zener interaction regions (to avoid overcrowding the picture, only the extreme relevant avoided crossings between the $n_0$ and $(n_0+1)$ manifolds have been shown).}
\label{fig1}
\end{figure}

\begin{figure}[htbp]
\centering\epsfig{file=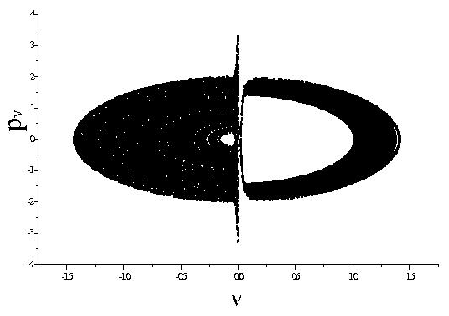,width=0.9\linewidth}
\caption{The Poincar\'e SOS in $v$ and $p_v$ for $n_0=40$ and $F n_0^5= 0.04$ (right) and $F n_0^5= 0.32$ (left). As the SOS is symmetric for $v \leftrightarrow -v$, only half of each SOS is shown}
\label{fig2}
\end{figure}

\begin{figure}[htbp]
\centering\epsfig{file=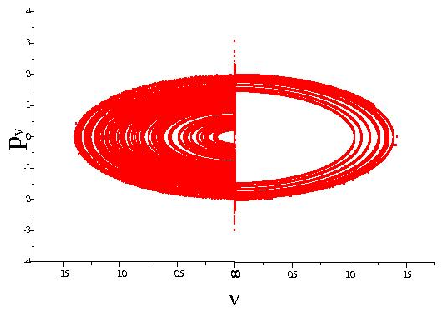,width=0.9\linewidth}
\caption{The Poincar\'e SOS in $v$ and $p_v$ for $n_0=320$ and $F n_0^5= 0.04$ (right) and $F n_0^5= 0.32$ (left). As the SOS is symmetric for $v \leftrightarrow -v$, only half of each SOS is shown}
\label{fig3}
\end{figure}

\begin{figure}[htbp]
\centering\epsfig{file=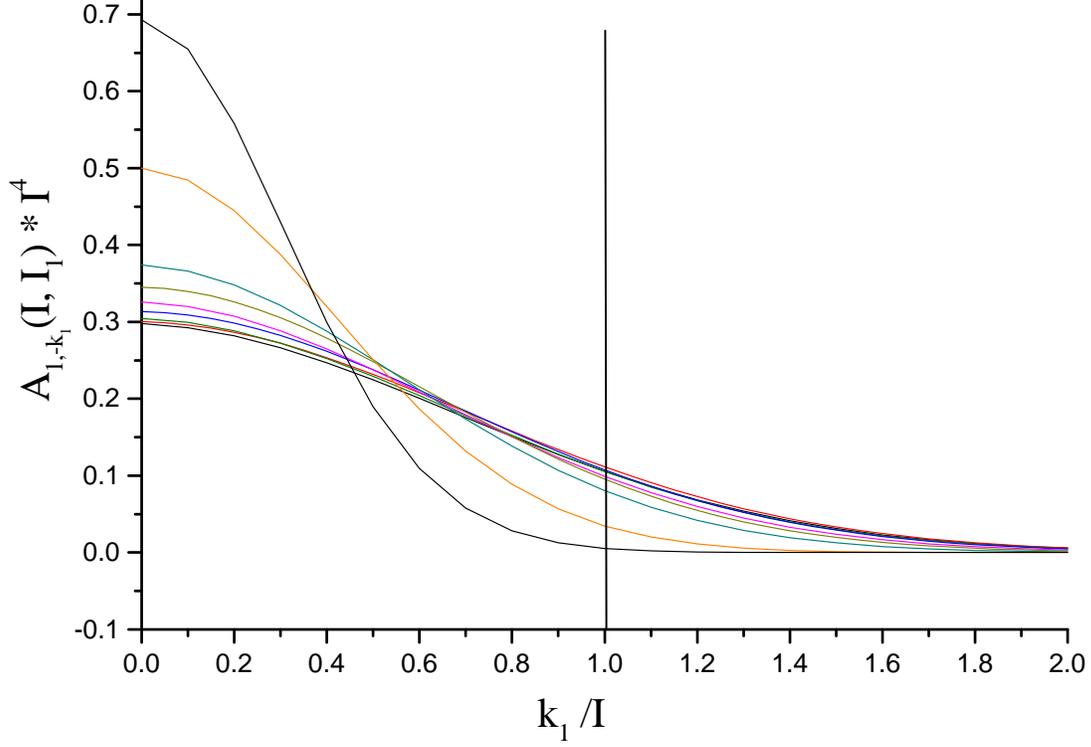,width=0.9\linewidth}
\caption{ (color online) The dependence of $\bar{A}_{1,-\bar{k}_1}(\bar{I}_1)$ on $\bar{k}_1$ for $\alpha=2.13$ and $\beta=2$ and for various values of $\bar{I}_1$. From top to bottom at $\bar{k}_1=0 $: $\bar{I}_1=0.05$, $0.10$, $0.20$, $0.25$, $0.30$, $0.35$, $0.40$, $0.45$, $0.50$.}
\label{fig4}
\end{figure}

\begin{figure}[htbp]
\centering\epsfig{file=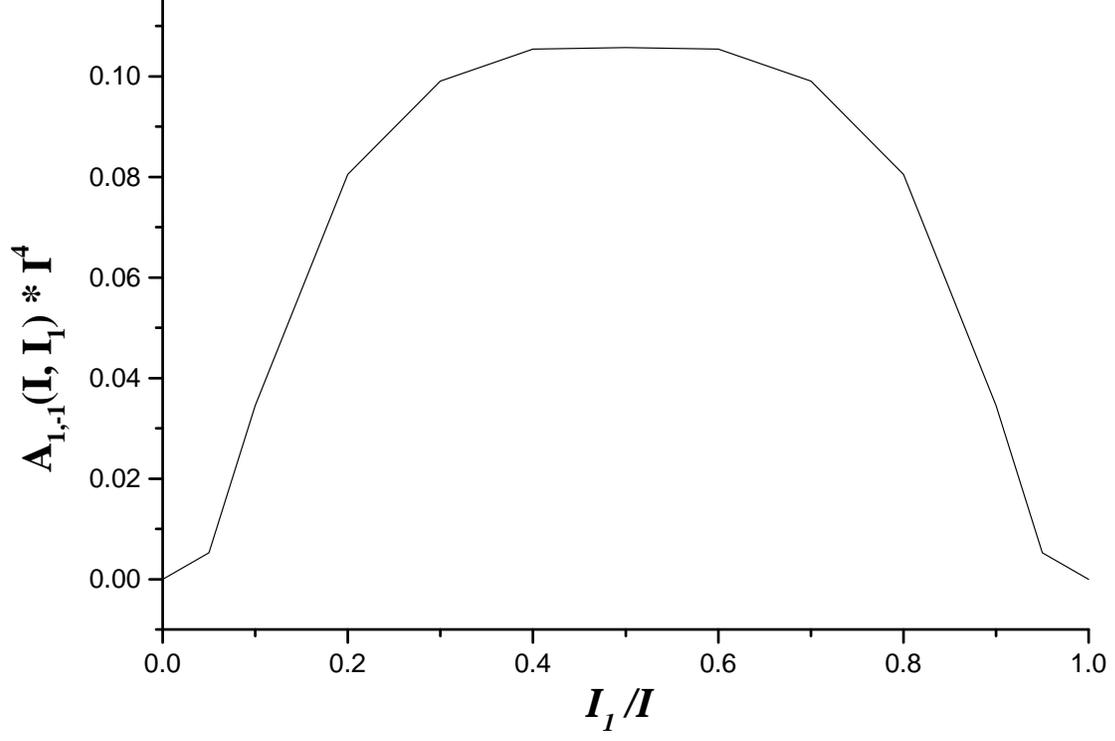,width=0.9\linewidth}
\caption{The dependence of $\bar{A}_{1,-1}(\bar{I}_1)$ on $\bar{I}_1$ for $\alpha=2.13$ and $\beta=2$.}
\label{fig6}
\end{figure}

\begin{figure}[htbp]
\centering\epsfig{file=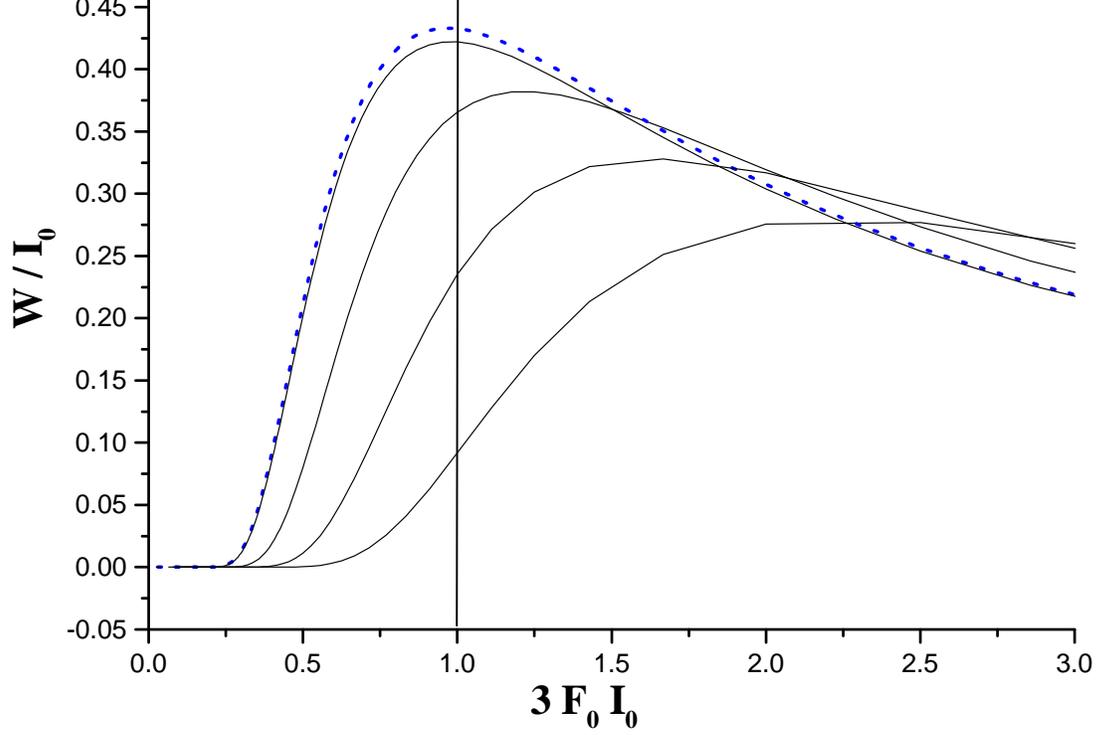,width=0.9\linewidth}
\caption{The dependence of $W/I_0$ on $3F_0 I_0$ for various values of $\bar{I}_1$: from top to bottom $\bar{I}_1=0.50$, $0.20$, $0.10$, and $0.05$. the dotted line is the approximated width calculated using eq. (\ref{wid}).}
\label{fig7}
\end{figure}

\begin{figure}[htbp]
\centering\epsfig{file=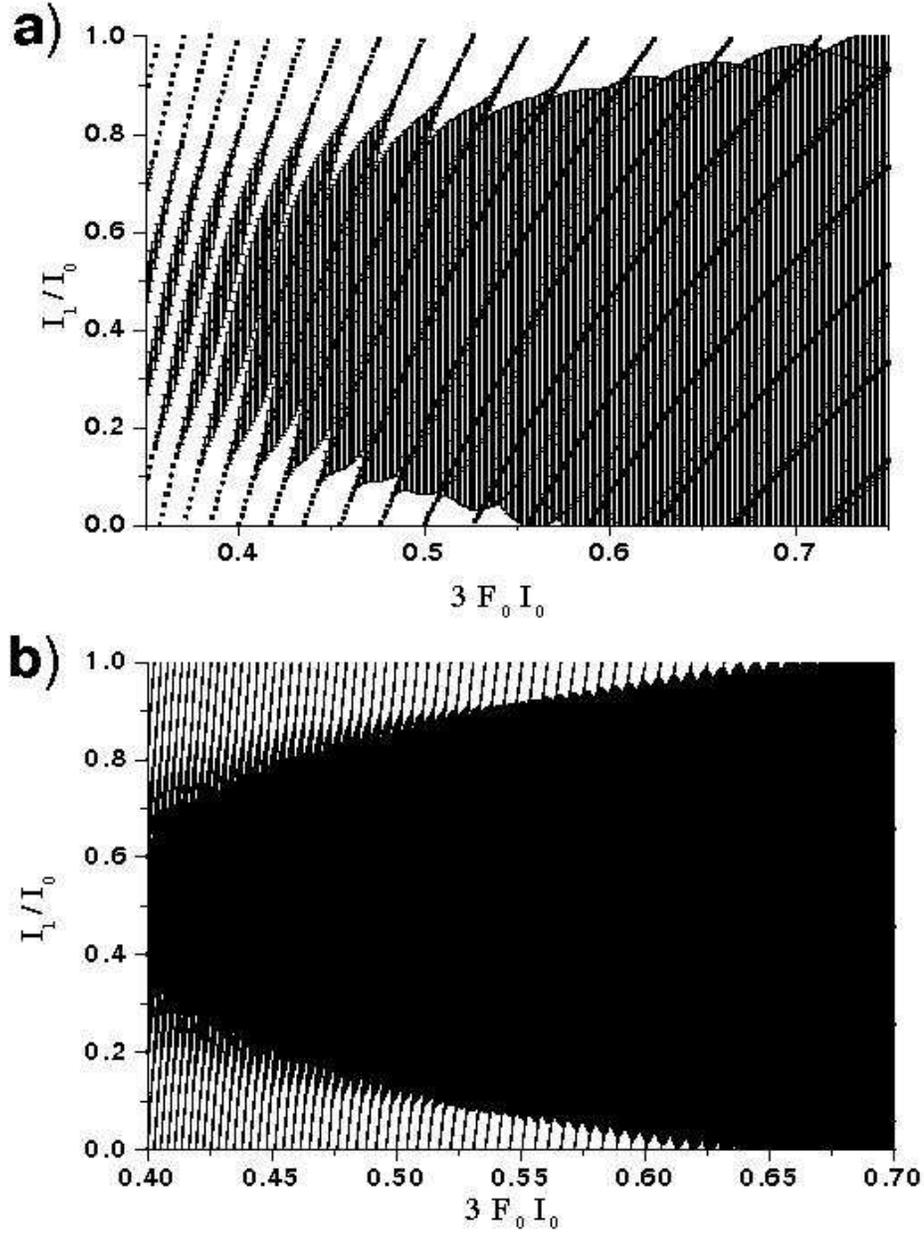,width=0.9\linewidth}
\caption{Position (eq. (\ref{n1})) and width (eq. (\ref{wid2})) of the resonances in $\bar{I}_1=I_1/I\simeq I_1/I_0$ as a function of $3F_0I_0$ for a) $I_0=10$; b) $I_0=60$.}
\label{fig8}
\end{figure}

\begin{figure}[htbp]
\centering\epsfig{file=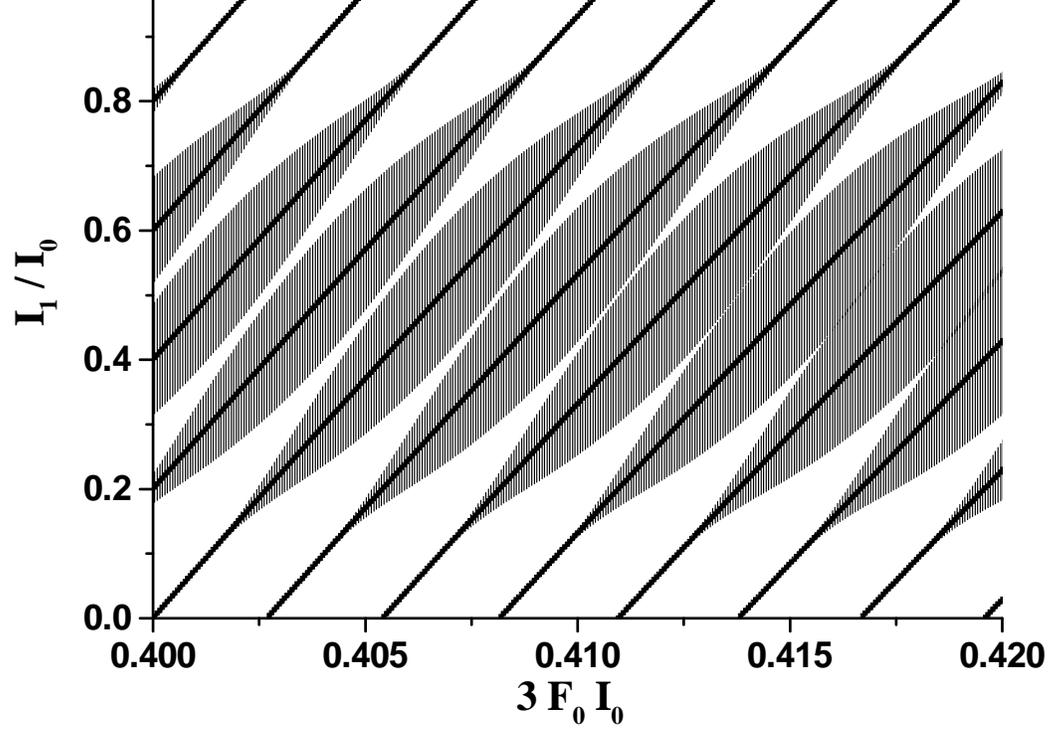,width=0.9\linewidth}
\caption{Detail of Fig. (\ref{fig8})b for the values of $3F_0I_0$ where the resonances first overlap.}
\label{fig9}
\end{figure}

\begin{figure}[htbp]
\centering\epsfig{file=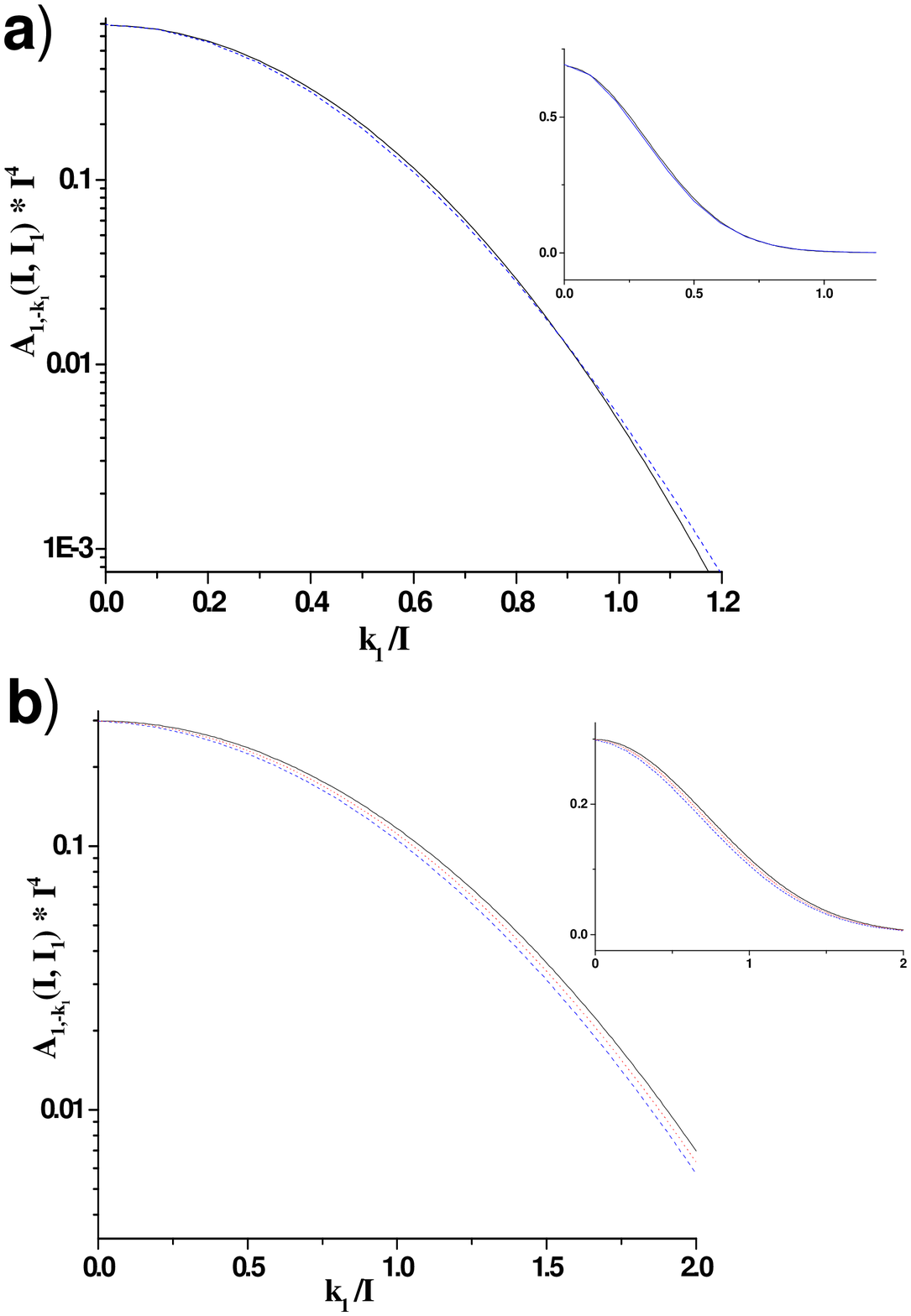,width=0.9\linewidth}
\caption{ (color online) Comparison of the exact dependence of $\bar{A}_{1,-\bar{k}_1}(\bar{I}_1)$ on $\bar{k}_1$ with the approximation using eq.(\ref{aaa2}). a) $\bar{I}_1=0.05$, full curve eq.(\ref{ii}), dash: exact result for $\bar{I}=10$. b) $\bar{I}_1=0.5$, full curve: eq.(\ref{ii}), dash: exact result for $\bar{I}=10$, dot: exact result for $\bar{I}=20$. the parameters are those for lithium: $\alpha=2.13$ and $\beta=2$. The main figures are on a logaritmic scale, the insets on a linear one.}
\label{fig5}
\end{figure}


\begin{thebibliography}{28}
\bibitem{note5} The quadratic term of the Stark shift reads 
\begin{equation}
U^{(2)} = -{7 \over 8} F^2 I^6 \left[{1+{6 \over 7} {{I_1} \over I}\left({1-{{I_1} \over I}}\right)+{{19}\over{14 I^2}}}\right].
\end{equation} 
Its ratio to the first order term is therefore of the order of $F I^4$, which for the fields of interest $F \sim < 1/(3I^5)$ is less than $1/(3I)$, giving a negligible contribution for $I \gg 1$.
\bibitem{note3} We note that in this case we have $F(t)=0$ and we therefore need not worry about it in the classical picture; as opposed to the quantum picture outlined in section \ref{due} where the non-adiabaticity at $F(t)=0$ plays a fondamental role.
\bibitem{note2} Even in resonance approximation classical adiabaticity in that region is not possible, as the motion on the separatrix has an infinite period, due to the presence of unstable fixed points.
\bibitem{note4}From eq. (\ref{kap}), the maximum speed of motion of the center of of a resonance in $I_1$ is 
\begin{eqnarray} 
{{{\dot I}_1^{max}} \over I}= {{\omega_f}\over {3F_0}}.\nonumber
\end{eqnarray}
\end{thebibliography}
\end{document}